# Erythrocyte-Inspired Discoidal Polymeric Nanoconstructs carrying Tissue Plasminogen Activator for the Enhanced Lysis of Blood Clots

Marianna Colasuonno[1,6], Anna Lisa Palange[6], Rachida Aid[3], Miguel Ferreira[6], Hilaria Mollica[2,6], Roberto Palomba[6], Michele Emdin[1,4], Massimo Del Sette[5], Cédric Chauvierre[3], Didier Letourneur[3], Paolo Decuzzi[6] *

[1] Sant'Anna School of Advanced Studies, Piazza Martiri della Libertà, 33, 56127, Pisa, Italy

[2] Department of Informatics, Bioengineering, Robotics and System Engineering, University of Genoa, Via Opera Pia, 13 Genoa 16145, Italy

[3] INSERM U1148, Laboratory for Vascular Translational Science, University Paris 13, University Paris Diderot, X. Bichat Hospital, 46 rue Henri Huchard, 75018, Paris, France

[4] Fondazione Toscana G. monasterio, Via G. Moruzzi, 1, 56124, Pisa, Italy

[5] S.C. Neurologia, E.O. Ospedali Galliera. Mura delle Cappuccine, 14. 16128 Genova

[6] Laboratory of Nanotechnology for Precision Medicine, Fondazione Istituto Italiano di Tecnologia, Via Morego, 30, 16163, Genoa

*  **Corresponding author**: Paolo Decuzzi, PhD. paolo.decuzzi@iit.it

**ABSTRACT**

Tissue plasminogen activator (tPA) is the sole approved therapeutic molecule for the treatment of acute ischemic stroke. Yet, only a small percentage of patients could benefit from this life-saving treatment because of medical contraindications and severe side effects, including brain hemorrhage, associated with delayed administration. Here, a nano therapeutic agent is realized by directly associating the clinical formulation of tPA to the porous structure of soft discoidal polymeric nanoconstructs (tPA-DPNs). The porous matrix of DPNs protects tPA from rapid degradation, allowing tPA-DPNs to preserve over 70 % of the tPA original activity after 3 h of exposure to serum proteins. Under dynamic conditions, tPA-DPNs dissolve clots more efficiently than free tPA, as demonstrated in a microfluidic chip where clots are formed mimicking *in vivo* conditions. At 60 min post treatment initiation, the clot area reduces by half (57 $\pm$ 8 %) with tPA-DPNs, whereas a similar result (56 $\pm$ 21 %) is obtained only after 90 min for free tPA. In murine mesentery venules, the intravenous administration of 2.5 mg/kg of tPA-DPNs resolves almost 90 % of the blood clots, whereas a similar dose of free tPA successfully recanalize only about 40 % of the treated vessels. At about 1/10 of the clinical dose (1.0 mg/kg), tPA-DPNs still effectively dissolve 70 % of the clots, whereas free tPA works efficiently only on 16 % of the vessels. *In vivo*, discoidal tPA-DPNs outperform the lytic activity of 200 nm spherical tPA-coated nanoconstructs in terms of both percentage of successful recanalization events and clot area reduction. The conjugation of tPA with preserved lytic activity, the deformability and blood circulating time of DPNs together with the faster blood clot dissolution would make tPA-DPNs a promising nanotool for enhancing both potency and safety of thrombolytic therapies.

In thrombolytic therapies, the rapid recanalization of occluded blood vessels is a key requirement to prevent extensive cell death and permanent tissue damage, and subsequent organ dysfunction and impair patient prognosis. Currently, major thrombotic events, such as ischemic stroke, myocardial infarction, pulmonary embolism, and venous thrombosis, are mostly treated with the systemic or local infusion of clot busting drugs.[1-3] Recombinant tissue plasminogen activator (tPA) is the most effective and sole clinically approved of these drugs. This protein catalyzes the conversion of blood plasminogen into plasmin, which eventually breaks down the fibrin network holding together the cellular component of the clot.[4] However, this same protein is known to induce bleeding and modulate the permeability of the blood brain barrier leading to intracranial hemorrhages. For this reason, clinical protocols suggest the slow intravenous infusion of tPA for up to 2 hours, and only within the first 3 to 4.5 hours after symptom's onset.[5] Delay administrations are associated with a higher risk of developing intracranial hemorrhages, neuronal damage and permanent disabilities, and fatal outcome, as a consequence of the progressive non-specific deposition of tPA within the brain parenchyma.[6] All these safety-related restrictions significantly limit the number of patients that could benefit of tPA-based therapies. In the case of acute ischemic strokes, it is estimated that less than 5 % of the patients would receive a tPA treatment and, out of them, 60 % would either suffer permanent disabilities or die.[7, 8]

A more efficient clinical exploitation of thrombolytic therapies would require the administration of agents with a high affinity with blood clots and negligible deposition within the brain parenchyma in order to realize selective thrombolysis with no hemorrhagic complications. Three different strategies have been proposed to improve the specific delivery of clot busting agents to vascular occlusions: conjugation with ligand molecules; loading into nanoparticles; and association with erythrocytes.

Within the first strategy, Haber and colleagues conjugated tPA molecules to an anti-fibrin monoclonal antibody and achieved successful fibrinolysis in a rabbit thrombosis model with an overall 3-fold increase in potency, as compared to free tPA.[9] Collen *et al*. developed a series of antibodies against activated platelets which were used *in vitro* and *in vivo* for treating platelet rich clots.[10] The group of Yang devised a modular approach for originating thrombolytic compounds resulting from the electrostatic interaction between a negatively charged heparin–antifibrin complex and a modified tPA molecule.[11, 12] Although these specifically targeted molecular agents have been demonstrated over 15 years ago, their clinical application has been limited, likely, for their complexity, stability, immunogenicity, and reduced thrombolytic activity as compared to the original free active molecule.[13]

More recently, nanoparticles have been proposed as an alternative strategy. After the pioneering work of Heeremans and colleagues, who loaded tPA molecules in the aqueous core of conventional liposomes showing a 5-fold increase in thrombolytic potency,[14] a variety of nanoparticles with clot-specific targeting ability and triggered release of thrombolytic agents have been demonstrated.[15-17] For instance, the group of Lanza and Wickline prepared fibrin-targeted liquid perfluorocarbon nanoparticles for the delivery of streptokinase demonstrating *in vitro* enhanced thrombolysis upon ultrasound stimulation.[15] Vaidya *et al*. developed target-sensitive liposomes that would rapidly release tPA in the presence of activated platelets, thus realizing a selective drug release.[16] Pegylated liposomes were used to increase up to 20-times the half-life of tPA in the circulation.[17] Recently, the group of Letourneur demonstrated that fucoidan-labeled polymeric nanoparticles exhibit a strong tropism for the clot fibrin network and facilitate vessel recanalization.[18] Non-organic nanoparticles have also been used for thrombolysis. Clusters of iron oxide nanocubes, coated and stabilized by a tPA multi-layer, were proposed by the authors for the combined chemical and thermal treatment of blood clots upon stimulation with exogenous alternating magnetic fields.[19] Ultrasmall superparamagnetic iron oxide nanoparticles, coated with fucoidan, were proposed for the magnetic resonance imaging (MRI) of blood clots.[20] The team led by Kim developed fibrin-targeted glycol chitosan–coated gold nanoparticles for assessing the clot stability *via* computed tomography (CT) imaging.[21] The advantage of nanoparticles over the free or targeted tPA stays in the larger amounts of lytic agents that could be deployed at the clot site; the multivalent adhesive interactions with the clot surface; the higher stability and longer circulation half-life *in vivo*; the ability to trigger the release *via* exogenous and endogenous stimuli, and to image and treat clots simultaneously.

Micron-sized particles and blood cells have been also proposed for the specific delivery of thrombolytic molecules to blood clots. A notable example is given by the work of Korin and colleagues who developed ∼ 4 μm spherical polymeric microparticles resulting from the aggregation of multiple, smaller ∼ 200 nm tPA-carrying nanoparticles. The microscale aggregates were shown to be intact under physiological flow conditions and break apart under high shear stresses (> 100 dyne/cm$^2$), which are typical of partially occluded vessels.[22] Although these shear-activated particles induced a 3-fold faster vessel recanalization in abdominal clots and rescued almost all tested mice with pulmonary emboli, they exhibited a short circulation half-life (< 5 min), which is comparable to free tPA, and a rapid accumulation in the liver (> 70 % ID/g). At even a larger scale, erythrocytes (RBCs) were elegantly used by the group of Muzykantov to deliver thrombolytic agents to nascent clots.[23-25] These blood cells have a characteristic size of ∼ 7 μm but their deformability allows them to squeeze through the smallest capillaries in the lungs and splenic parenchyma, supporting their circulation for several days. Indeed, tPA-RBCs cannot infiltrate or firmly attach onto mature clots, because of their large size. However, they can certainly impair the formation of new clots and, as such, were proposed for thromboprophylaxis rather than for acute treatments.[23] Moreover, the large size of tPA-RBCs prevented the accumulation of free tPA within the brain tissue, dramatically reducing the risk of cerebral hemorrhages.[25] Indeed, a major technical challenge for the clinical implementation of this approach is the coupling of tPA with autologous RBCs and their reinjection into the patient.

In this work, tPA molecules are conjugated to the polymeric structure of soft, long circulating discoidal nanoconstructs (tPA-DPNs), which would mimic erythrocytes.[26] These polymeric nanoconstructs are made out of a mixture of poly(lactic-co-glycolic acid) (PLGA) and polyethylene glycol (PEG) and present a diameter of 1,000 nm and a height of 400 nm. tPA molecules are covalently linked to the carboxyl groups exposed on the PLGA chains. In the sequel, the synthesis and stability of tPA-DPNs is first discussed, followed by the *in vitro* characterization of their clot dissolution potency. This is performed under static conditions and in an *ad-hoc* two-channel microfluidic chip where whole blood clots are generated and exposed to a tPA-DPN solution under continuous flow (dynamic conditions). Finally, the *in vivo* efficacy of tPA-DPNs is assessed in a mouse thrombosis model where blood clots are induced in mesentery venules. For all experiments, the lytic potential of tPA-DPNs is directly compared to free tPA and 200 nm spherical polymeric nanoconstructs (tPA-SPNs), made of PLGA and decorated on the surface with tPA molecules.

**RESULTS**

**Synthesis and Characterization of tPA-DPNs.** The synthesis of DPNs is a multistep, top-down process.[26-28] Briefly, laser writing lithography is used to realize a pattern with billions of wells, reproducing the characteristic geometry of DPNs, in a silicon master template (**Supporting Figure.1A**). Then, *via* replica molding, this template is used to generate an intermediate PDMS template (**Supporting Figure.1B**) and eventually a sacrificial PVA template (**Supporting Figure.1C**). In the current configuration, DPNs are circular disks with a diameter of 1,000 nm, a height of 400 nm, and are composed by an homogeneous mixture of Poly(D,L-lactide-co-glycolide)-acid carboxylic terminated (PLGA-COOH) and poly(ethylene glycol) diacrylate (PEG-DA) (**Figure.1A**). Note that these polymers are among the most studied and best characterized polymers for biomedical applications, and have been approved by various regulatory agencies for diverse clinical applications.[29, 30]

This polymeric mixture, together with a photoinitiator, is deposited into the wells of the PVA template and exposed to a UV-light source for crosslinking. After polymerization, the sacrificial PVA template is dissolved into an aqueous solution at room temperature, under gentle stirring, and, upon centrifugation and filtration, DPNs are collected. Note that DPNs are hydrogel-based particles with a porous structure. Finally tPA is directly associated to the resuspended DPNs. Specifically, the carboxylic groups on the PLGA chains are activated, *via* an EDC/NHS reaction, and then covalently coupled to one of the amine groups on the tPA molecules (**Supporting Figure.1D**). A schematic representation of tPA-DPNs is depicted in **Figure.1A**.

The geometrical features of DPNs and tPA-DPNs were assessed *via* a Multisizer Particle Counter, a Zetasizer Nano, electron microscopy, and atomic force microscopy (**Figure.1B-E**). The Multisizer Particle Counter spectra for the DPNs and tPA-DPNs are almost perfectly overlapped, as shown in **Figure.1B** (orange: tPA-DPNs; green: empty DPNs). After the reaction with tPA, DPNs presented a negligible increase in average size (< 3 %), from 784 $\pm$ 10 nm (DPNs) to 803 $\pm$ 15 nm (tPA-DPNs). On the other hand, the change in surface $\zeta$ potential was more relevant, increasing from a negative value of -21 $\pm$ 0.6 mV for DPNs to a positive value of 16 $\pm$ 0.4 mV for tPA-DPNs. This large variation in surface $\zeta$ potential has to be ascribed to the tPA conjugation with the carboxylic groups presented on the PLGA chains and the progressive neutralization of the originally negative surface electrostatic charge of DPNs. As such, the change in surface ζ potential can be also used as an indirect method to quantify the efficiency of tPA conjugation. A representative Transmission electron Microscopy (TEM) image of DPNs is given in **Figure.1C**, which confirms the discoidal shape with a circular base of about 1 μm in diameter. In the same figure, the upper-left inset shows an image deriving from the superimposition of an electron microscopy and a fluorescent microscopy picture for the same DPN, which was loaded with lipid-Rhodamine B (RhB). A Scanning Electron Microscopy (SEM) image and an Atomic Force Microscopy (AFM) image of DPNs are given in **Figure.1D-E**, respectively, which document a DPN diameter of ∼ 1.2 μm, for both SEM and AFM, and height of ∼ 347 nm for SEM and ∼ 300 nm for AFM. Finally, as depicted in **Figure.1F**, tPA-DPNs were demonstrated to be geometrically stable, upon storage in saline at 37° C, with an overall variation in diameter of 6 % over six days.

The pharmacological properties of DPNs were characterized for the tPA encapsulation efficiency and release profile (**Figure.2**). First, to confirm the firm conjugation of the thrombolytic agent to the particle structure, DPNs were grafted with FITC-tPA and loaded with lipid-RhB. This fluorescent lipid was dispersed uniformly within the PLGA/PEG matrix before particle synthesis. In **Figure.2A**, projections of tridimensional confocal microscopy images show the green fluorescence associated with FITC-tPA properly co-localized with the red fluorescence given by the lipid-RhB. The full reconstruction is presented in the **Supporting Figure.2**. This also demonstrates the ability to load multiple agents within the tPA-DPN matrix. Moreover, in order to assess the tPA encapsulation efficiency, $5 \times 10^8$ DPNs were incubated with different amounts of tPA, ranging from 10 to 70 μg. As depicted in **Figure.2B**, the amount of tPA conjugated to DPNs grows linearly with the input dose returning an almost 1:1 ratio (dashed line), thus implying an encapsulation efficiency close to 100 %. For $5 \times 10^8$ DPNs, saturation is reached at about 60 μg of tPA. Indeed, the total tPA amount can be finely tuned by changing the number of DPNs. Given the hydrogel nature of these DPNs, tPA molecules are expected to react with the PLGA-COOH groups exposed on the surface as well as those available within the polymer matrix. Finally, the possible release of tPA from DPNs was analyzed upon particle incubation in PBS, at 37° C, up to 72 h. **Figure.2C** shows that the largest majority of tPA (*i.e.* > 90%) is associated with the DPN structure throughout the observation period, which demonstrates the stable conjugation of the lytic agent. It is here important to highlight that this feature of tPA-DPNs together with their relatively large size would limit the non-specific accumulation of tPA molecules within the brain parenchyma and possibly reduce the risk of hemorrhagic events.

To assess if the conjugation of tPA to the DPNs surface affects the activity of the drug, an enzymatic assay of free and conjugated tPA was performed. This *in vitro* assay measures the ability of tPA to activate the plasminogen to plasmin, using a plasmin substrate releasing a yellow para-nitroaniline (pNA) chromophore. Three different concentrations were tested, namely 10, 20, and 50 µg/ml of drug, up to 2 h. As depicted in **Figure.2D**, the activity of tPA after the conjugation with the carboxylic group of PLGA is retained. No statistically significant difference is observed between free or bound tPA.

For comparison, Spherical polymeric nanoconstructs (SPNs) were also prepared. These nanoparticles comprise a PLGA core and a lipid surface layer, and are synthesized *via* an emulsion technique, as described in the **Supporting Information**. Similarly to tPA-DPNs, carboxylic groups exposed on the SPN surface are activated, *via* an EDC/NHS reaction, and then covalently coupled to the amine groups on the tPA molecules returning the tPA-SPNs. These conventional nanoconstructs have a spherical shape with a diameter of 240 ± 0.2 nm and a surface potential $\zeta$ = - 8 ± 0.6 mV. Dynamic light scattering data and scanning electron microscopy images are provided in **Supporting Figure.3**.

***In vitro* efficacy of tPA-DPNs under static conditions.** After characterizing the physico-chemical properties of tPA-loaded nanoparticles, their dissolution potential was assessed on fresh blood clots. First, a static experiment was performed in order to confirm that tPA would preserve its thrombolytic activity even after direct conjugation. Whole blood was collected from rats and immediately incubated with a thrombin solution to induce rapid clot formation. After maturation, clots were transferred in a 24-multi-well dish and treated with PBS (control), free tPA, SPNs, tPA-SPNs, DPNs, and tPA-DPNs. The same dose of tPA (30 µg/ml) was used for all conditions. **Figure.3A** shows representative images of treated blood clots at 5 selected time points, namely 0, 30, 90, 180 and 300 min. The progressive red color of the wells is associated with the break down of the fibrin network and release of the red blood cells. Notably, this happens only for free tPA, tPA-SPNs and tPA-DPNs solutions, demonstrating that the clots are otherwise stable. At these specific time points, the thrombolytic efficacy was quantified by measuring the optical density (OD 415) of the supernatant for each sample, related to clot lysis. As depicted in **Figure.3B**, blood clots treated with DPNs and SPNs were stable as the ones treated with PBS (CTR). This demonstrates that DPNs alone do not induce any damage to the red blood cells trapped within the fibrin network. In both control cases, a negligible dissolution of the clots is observed over time, which is related to the natural and spontaneous break down of the fibrin network. For the free tPA, tPA-SPNs and tPA-DPNs, blood clots were dissolved at much faster rates and with a comparable efficiency. At later time points, no statistically significant difference was observed between tPA, tPA-DPNs, and tPA-SPNs values reported in **Figure.3B** (p-values listed in the Supporting Information). Dissolution rates, defined as the ratio of the difference in optical density between two consecutive time points to overall time, are presented in **Figure.3C**. The data confirmed the faster dissolution of the clots when treated with the thrombolytic solutions (tPA, tPA-SPNs, and tPA-DPNs) and demonstrate that under *in vitro* quiescent conditions there is no difference in therapeutic efficacy between free tPA, tPA-SPNs, and tPA-DPNs. More importantly, it is confirmed that the direct conjugation of tPA to the polymeric structures of SPNs and DPNs does not affect the lytic potential of tPA.

Data in **Figures.3A-C** are related to a tPA dose of 30 µg/ml per clot. Similar experiments are presented in the **Supporting Information** for lower (10 µg/ml) and higher (50 µg/ml) doses of tPA (respectively, **Supporting Figures.4A-D**). As expected, the efficacy of clot dissolution grows with the tPA concentration. In particular, for a dose of 10 µg/ml, the dissolution rate is found to be 0.01 $min^{-1}$ and grows to 0.02 and 0.03 $min^{-1}$ for the 30 and 50 µg/ml cases, respectively. Once again, this data confirm that the direct conjugation of tPA onto SPNs and DPNs does not impair the thrombolytic activity of the clinical agent. **Figure.3D** shows two electron microscopy images of blood clots before and after exposure to tPA-DPNs. In the first image, blood cells form a dense matrix glued together by long and abundant fibrin chains. After treatment, the fibrin network has almost disappeared and the clot cell density clearly diminished.

Until now, all dissolution tests were performed in PBS which lacks blood proteins such as albumin, hormones, and others macromolecules that could adsorb on the particle surface and interfere with the lytic process or even directly affect the stability of tPA. Therefore, dissolution experiments were also performed in the presence of 100 % fetal bovine serum (FBS) and data are reported in **Figure.4**. For the three different experimental groups (FBS, free tPA and tPA-DPNs), representative images of the clots at different time points (**Figure.4A**), optical density quantifications (**Figure.4B**), and dissolution rates (**Figure.4C**) are presented. Note that in the presence of serum, the same trends as with PBS experiments were obtained. FBS has a mild absorbance at 415 nm, which would explain the not-zero dissolution values observed over time for the control group (FBS alone). Indeed, if this contribution is subtracted from the optical density of tPA and tPA-DPNs, the results obtained in FBS are comparable with those in PBS.

This is a preliminary demonstration that the lytic activity of tPA-DPNs is not affected by presence of serum proteins. Moreover, the lytic stability of tPA-DPNs and tPA-SPNs was assessed upon incubation of the nanoconstructs with FBS. Specifically, tPA-DPNs and tPA-SPNs were first incubated for 0.5, 1, and 3 h with FBS and, then, tested for their blood clot dissolution activity. As expected, pre-incubation of tPA-DPNs and tPA-SPNs with FBS reduces their lytic activity (**Figure.4D**). Interestingly, the lytic activity reduces progressively with the pre-incubation time only in the case of tPA-DPNs. Unexpectedly, even after 3 h of incubation with serum proteins in FBS, over 70 % of the thrombolytic activity of tPA-DPNs is retained, as compared to fresh free tPA. This is an interesting result, given that free tPA has a very short stability in blood. In the case of tPA-SPNs, a 50 % reduction in lytic activity is registered within the first 30 min of incubation followed by a steady slow decrease for longer incubations. Indeed, thrombolytic molecules are directly exposed on the nanoparticle surface in tPA-SPNs and immediately accessible by blood proteins. In contrast, the thrombolytic molecule is distributed within the hydrogel porous matrix of the nanoconstructs in tPA-DPNs. These results would suggest that the surface adsorption of blood proteins, which happens within the first minute following FBS exposure, is not significantly affecting the activity of tPA associated with DPNs.

***In vitro*** **efficacy of tPA-DPNs under dynamic conditions.** The thrombolytic activity of tPA-DPNs was also evaluated under dynamic conditions. For the formation and dissolution of blood clots, a microfluidic chip was fabricated consisting of two parallel channels separated in the middle by a row of pillars (**Supporting Figure.4A**).[31] Whole blood was flowed in the bottom channel while an aqueous solution of thrombin with the addition of $CaCl_2$, was infused in the upper channel (**Figure.5A, left**). Since whole blood was collected in tubes containing sodium citrate, the $CaCl_2$ added to the thrombin solution was used to reactivate the process only within the microfluidic chip. The infusion of the two solutions was controlled *via* a syringe pump. The pro-coagulant solution encounters whole blood in the middle of the chip where the row of pillars facilitates their mixing and the spontaneous formation of the blood clot, as depicted in **Figure.5A** (**right**) and in the **Supporting Movie.1**.

The so-formed clots were treated with PBS, SPNs, DPNs, free tPA, tPA-SPNs, and tPA-DPNs, using the same dose as per the static experiments (*i.e.* 30 µg/ml). After clot formation, thrombolytic solutions were infused in both the upper and lower channels at 100 nl/min, for flow conditions comparable to that of human arterioles. The fluidic experiments were performed in a micro-environmental chamber placed on the stage of a time-lapse microscope, preserving a constant temperature at 37° C and clots were monitored for 2 h continuously. In contrast to static experiments, clots form spontaneously within the microfluidic chips and, similarly to *in vivo* experiments, their initial size cannot be precisely controlled. As such, for the quantitative analyses, areas of the clots were weighted-averaged with respect to their initial size for each experimental group. The results indicated that empty SPNs and DPNs did not affect significant the fibrin network around the clot exhibiting a behavior similar to that of PBS. A modest dissolution of the clot is documented within the first 30 min for both empty SPNs and DPNs, which should mostly be associated to the intrinsic stability and progressive maturation of the blood clot. Movies and images for these cases are shown in **Supporting Figures.6B-C** and **Supporting Movies.2-3**. In contrast, free tPA, tPA-SPNs and tPA-DPNs were capable of dissolving the blood clots, as documented by the representative images of **Figure.5B**, for free tPA, and **Figure.5C**, for tPA-DPNs. The full clot dynamics is provided in the **Supporting Movies.4-5**. The variation of the clot size over time, weighted-averaged for its initial size, is shown in **Figure.5D**, for all experimental groups. The clot size significantly reduces with time for the thrombolytic solutions (tPA, tPA-SPNs and tPA-DPNs), while no appreciable change is observed for the control groups (PBS, SPNs and DPNs). Importantly, the two tPA nanoconstructs (tPA-SPNs and tPA-DPNs) show a more efficient dissolution of the blood clots as compared to free tPA. Already at 60 min post treatment initiation, the clot size reduces down to $57 \pm 8$ % for tPA-DPNs and $42 \pm 10$ % for tPA-SPNs against $83 \pm 10$ % for free tPA. These numbers decrease even more at 90 and 120 min, being respectively $34 \pm 7$ % and $28 \pm 12$ % for the tPA-DPNs and $24 \pm 3$ % and $17 \pm 6$ % for the tPA-SPNs against $56 \pm 21$ % and $44 \pm 19$ % for free tPA (**Figure.5D**). As depicted in **Figure.5E**, the dissolution rates of tPA-SPNs and tPA-DPNs are significantly higher than for free tPA. At 60 min post treatment initiation, the dissolution rate of tPA-DPNs is over 2-fold higher than for free tPA and 50% higher than for tPA-SPNs. At 30 min, tPA-SPNs present a dissolution rate which is two times larger than for tPA-DPNs. Both **Figure.5D** and **E** document a faster clot dissolution within the first 60 to 90 min post-treatment initiation for tPA-DPNs over free tPA. This is indeed a key feature of this nano-thrombolytic agent as compared to the molecular tPA.

The higher efficacy observed for tPA-DPNs over free tPA should be ascribed to the fact that DPNs are large enough and deformable to efficiently deposit on the clot surface and be entrapped within the fibrin network. This would increase the residence time of tPA in the clot proximity and thus enhance the lytic activity of the nano-agent. On the other hand, the molecular form of tPA would be rapidly washed away, thus providing a limited the temporal interaction with the blood clot. This was confirmed by infusing within the microfluidic chip RhB-DPNs and monitoring their interaction dynamics with blood clots with an epifluorescent microscope. The **Supporting Movie.6** demonstrates the progressive accumulation of the RhB-DPNs (red labeled) on and around the blood clot. Although this movie confirms the ability of DPNs to passively interact with the clot network, it should be also kept in mind that tPA has a direct affinity for fibrin.[9] Moreover, discoidal particles have been shown in multiple studies to enhance adhesion under flow.[32-34] Thus, the observed accumulation of tPA-DPNs at the clot site could also result from the appropriate combination of tPA-mediated specific adhesion with DPN geometry. On the other hand, the faster dissolution rates associated with tPA-SPNs at earlier time points should be ascribed to the ability of these smaller particles to infiltrate the fibrin network of the clot and the direct exposure of the tPA molecules on the SPN surface.

***In vivo* efficacy of tPA-DPNs.** A thrombosis model was established in C57BL/6 mice by inducing a localized vascular injury. Specifically, a filter paper was soaked in a $FeCl_3$ solution and then applied over the surface of an isolated venule in the murine mesentery circulation.[18, 19] In few seconds, a blood clot was locally generated as documented by intravital microscopy (**Supporting Movie.7**). After thrombus formation, 200 μl of treatment solutions were retro-orbitally injected. The temporal evolution of the blood clot was followed in response to seven treatment groups: DPNs, SPNs, 1 mg/kg free tPA, 2.5 mg/kg free tPA, 1 mg/kg tPA-SPNs, 2.5 mg/kg tPA-SPNs, 1 mg/kg tPA-DPNs, and 2.5 mg/kg tPA-DPNs. Platelets and leukocytes were labelled in red with an injection of Rhodamine 6G. The size and fluorescent intensity of the blood clots was monitored over time up to 35 min post treatment initiation (**Supporting Movies.8-9**). Based on the temporal evolution of the clots, these were divided in two groups: beneficial clots and critical clots (**Figures.6A-C**). Beneficial are those clots that present a reduction in size, in fluorescence signal, or both within the first 35 min of observation

(**Figure.6B**). Critical are those clots that do not present any significant reduction in size or fluorescence (**Figure.6C**).

Out of all tested animals (n ≥ 10 mice per experimental group), **Figure.6D** summarizes the occurrence of beneficial and critical clots for the five tested conditions. For DPNs and SPNs, all clots are critical (red bars). This result confirm that nanoconstructs alone do not cause any injury at the clot site. As the tPA concentration increases, the occurrence of beneficial clots increases as well. For 1 mg/kg of thrombolytic agent, only 16 % of the clots were resolved for free tPA, 10 % for tPA-SPNs and 60 % for tPA-DPNs. At 2.5 mg/kg of thrombolytic agent, the percentage of beneficial clots was 43 % for free tPA, 20 % for tPA-SPNs and 90 % for tPA-DPNs. The percentage variation of the clot area at 35 min post treatment initiation is depicted in **Figure.6E**. As expected, DPNs and SPNs did not cause any appreciable change in the clot area. The clot area reduced to 88 ± 3 % and 81 ± 6 % for 1.0 and 2.5 mg/kg free tPA, respectively. More significant reductions in clot area are documented for tPA-DPNs: for 1 mg/kg tPA-DPNs, the area reduces to 78 ± 3 %, and down to 58 ± 8 % for 2.5 mg/kg tPA-DPNs. Interestingly, tPA-SPNs induce a clot area reduction which is comparable to that of free tPA: 92 ± 2 % for 1 mg/kg tPA-SPNs and 86 ± 2 % for 2.5 mg/kg tPA-SPNs. The difference in area reduction is statistically significant for the two different tPA doses within all three groups (free tPA, tPA-SPNs and tPA-DPN) demonstrating that the increase in tPA dose from 1 to 2.5 mg/kg is effective. Also, the difference in area reduction is statistically significant between free tPA and tPA-DPNs at each of the two tPA doses, demonstrating that tPA conjugation to the DPN structures always enhances the thrombolytic activity of the clinical agent. Although, tPA-SPNs were capable of dissolving clots *in vitro* more rapidly than tPA-DPNs (**Figures.4E** and **D**), they did not show *in vivo* any significant advantage over free tPA. This should be ascribed to the more efficient circulation profiles of tPA-DPNs.

It is here important to highlight that in humans, the administered dose of tPA is of about 1 mg/kg, depending on the duration of the infusion.[35] Using an animal equivalent dose calculation based on the body area,[36] this would correspond to over 10 mg/kg in mice. Given the thrombolytic potency of tPA-DPNs even at 1 mg/kg of tPA, it can be inferred that these nano-thrombolytic agent would be at least 10-fold more effective than the clinical formulation.

DISCUSSION

From the above results, key distinctive features of tPA-DPNs over its free molecular form (tPA) and the nanometric spherical counterpart (tPA-SPNs) can be readily deduced: firstly, the efficient and stable association of tPA with the DPN polymeric structure; secondly, the size, shape and deformability combination favoring deposition within blood clots; and thirdly, the rapid dissolution of fibrin based vascular occlusions.

As shown in **Figures.2A-B**, DPNs can be associated with tPA returning a loading efficiency close to 100 %. This is achieved by using the available clinical formulation of tPA, with no additional modification or purification steps. Furthermore, the association is highly stable, as demonstrated in **Figure.2C**, whereby less than 10 % of the conjugated tPA would detach from the DPN structure and be freely released in the blood stream. This stable association together with the micrometric size of DPNs would limit their accumulation within the brain parenchyma, even in the presence of a hyperpermeable vascular bed. Indeed, this could be crucial in preventing the occurrence of the dramatic side effects associated with the systemic administration of tPA, in line with what has been documented by Muzykantov and colleagues for the large tPA-RBCs.[25] Notice that a similar behavior would not be expected for the 200 nm tPA-SPNs, which would likely extravasate with their tPA load at sites of vascular hyper-permeability.

In addition to the conjugation stability, the lytic activity and geometrical features of tPA-DPNs appears to be retained quite efficiently even after exposure to serum proteins. **Figure.1F** has documented that the incubation of DPNs in a physiological solution does not alter their size and shape configurations for several days. The surface zeta potential of tPA-DPNs in physiological solution reduces slowly with time (**Supporting Figure.5)** and the lytic activity of tPA-DPNs appeared to be well preserved over the first few hours (**Figure.4D)**. Even after 3 h post incubation with FBS, tPA-DPNs provide a lytic activity corresponding to about 70 % of that of fresh free tPA. Since tPA is associated with the carboxylic groups on the PLGA chains, which are distributed within the whole DPN matrix, tPA molecules would be attached on the DPN surface as well as distributed within the DPN matrix. This is demonstrated by the confocal images in **Figure.2A**, with FITC-tPA uniformly distributed across the entire DPN matrix. This could explain the retained lytic activity whereby the outer most tPA molecules would undergo degradation followed by the inner most tPA molecules. Indeed, this can also be inferred by analyzing the data in **Figure.4D** for the tPA-SPNs. In these spherical nanoconstructs, tPA molecules are all exposed on the surface and, consequently, the tPA-SPN lytic activity decreases abruptly to 30 % within the first 30 min. It is here important to note that the slightly positive surface charge of tPA-DPNs should be mostly ascribed to the neutralization of the surface PLGA-COOH groups and can be finely tuned by modulating tPA loading (**Supporting Figure.6**).

Another attribute of tPA-DPNs is their size, shape and deformability combination. The DPN geometry has been selected for its higher efficacy in adhering at surfaces under flow. This has been demonstrated, by the authors and other researchers, using in silico computational models, *in vitro* fluidic chips, and *in vivo* small animal experiments. [32-34, 37, 38] Briefly, the efficient adhesion of discoidal nanoparticles to blood clots would result from the fine balance between wall adhesive interactions, which are enhanced by the large discoidal surface, and the dislodging hydrodynamic forces, which are reduced by the thin cross section of the particles. Although, in the current configuration, tPA-DPNs do not carry any targeting moiety for blood clots, it should be recalled that tPA has a significant affinity for fibrin. Thus, specific adhesive interactions could arise at the interface between a tPA-DPN and the fibrin network in the clot. Importantly, this adhesive interaction would involve simultaneously multiple tPA molecules (multivalent interaction) exposed over the large discoidal DPN surface. Furthermore, differently from conventional nano-thrombolytic agents, tPA-DPNs are made using the so-called 'soft' version of DPNs. These particles have a Young's modulus ranging between a few tens to a few hundreds of kPa.[28] The deformability of tPA-DPNs would favor their deposition within the complex fibrin network of the blood clot, thus providing an additional mechanism of accumulation. Indeed, as previously shown by the authors, DPN deformability also allows them to navigate in the blood stream longer, returning a circulation half-life of about 22 h, and resist internalization by phagocytic cells and vascular endothelial cells (**Supporting Figure.7**).[26] In contrast, previous bioavailability and biodistribution studies on SPNs indicated a circulation half-life of few hours, in agreement with most 200 nm spherical particles documented in the literature.[39] This is also confirmed by a side-by-side comparison performed between Cy5.5-labeled DPNs and SPNs in immunocompetent mice (**Supporting Figure.8**), which shows a much higher accumulation of SPNs into major reticuloendothelial organs (liver, kidneys and lungs) as compared to DPNs.

All these features of tPA-DPNs result in an agent that can more rapidly and efficiently dissolve blood clots as compared to free tPA. Under dynamic conditions *in vitro* (**Figure.5**), tPA-DPNs reduce the clot size by almost 50 % already within the first 60 min of treatment, whereas free tPA would require over 90 min to achieve the same result. In the mouse mesentery vasculature (**Figure.6**), 2.5 mg/kg tPA-DPNs can recanalize almost 90 % of the occluded vessels exhibiting 50 % reduction in clot size after only 35 min post treatment initiation. The same dose of free tPA can recanalize less than half of the occluded vessels with about a 80 % reduction in clot size within the first 35 min.

**CONCLUSIONS**

A thrombolytic agent has been prepared by conjugating molecules of tissue plasminogen activator to the polymeric matrix of soft discoidal nanoconstructs – tPA-DPNs. Under static conditions, tPA-DPNs have the same lytic behavior as free tPA and 200 nm tPA-spherical PNs, both in terms of absolute dissolution and dissolution rates. This demonstrates that the direct reaction of the tPA ammine groups with the PLGA carboxylic groups in DPNs does not affect the original molecular activity of tPA. Furthermore, tPA distributes quite uniformly throughout the whole DPN porous matrix, with some tPA molecules directly exposed over the nanoconstruct surface whereas others would be embedded within them. Likely, this would explain the preserved lytic activity of tPA-DPNs upon exposure to serum proteins. Under dynamic conditions, in microfluidic chips, tPA-DPNs outperform free tPA molecules offering a faster degradation rate, which is indeed a crucial feature for any thrombolytic agent. In preclinical mouse experiments, tPA-DPNs were observed to dissolve clots more efficiently than free tPA at concentrations of both 1.0 and 2.5 mg/kg, which are 4 to 10 times lower than the clinical values. In contrast, no significant improvements over free tPA were documented *in vivo* for the 200 nm tPA-spherical PNs. In summary, the higher stability together with the more effective and faster dissolution of tPA-DPNs over free tPA, would make these constructs a promising nanotechnological platform for the treatment of acute thrombotic events.

**MATERIALS AND METHODS**

**Chemicals.** Polydimethylsiloxane (PDMS) (Sylgard 184) and elastomer were purchased from Dow Coming Corp (Midland, USA). Poly(vinylalcohol) (PVA, Mw 31,000 - 50,000), Poly(DL-lactide-co-glycolide) acid (PLGA, lactide:glycolide 50:50, Mw 38,000-54,000), Poly(ethylene glycol) diacrylate (Mn 750) (PEG diacrylate), 2-Hydroxy-40-(2-hydroxyethoxy)-2-methylpropiophenone (Photo-initiator), 1-Ethyl-3-(3-dimethylaminopropyl)-carbodiimide (EDC), N-Hydroxysuccinimide (NHS), FITC-tPA, and Thrombin were purchased from Sigma-Aldrich (Missouri, USA). Rhodamin-B (lipid-RhodB) was purchased from Avanti Polar Lipids (Alabama, USA). tPA was provided by Ospedale Galliera, (Liguria, IT). Bicinchoninic protein assay dye kit was purchased from BioRad (California, USA). All the reagents and other solvents were used without further purification.

**Synthesis of Discoidal Polymeric Nanoconstructs (DPNs) coated with tissue Plasminogen Activator (tPA).** Particles were synthesized using a top down approach previously described.[26, 28] Briefly, a silicon master template was fabricated using a Laser Writer Lithography technique which allows to transfer on the silicon a specific pattern made out of discoidal wells with a diameter and a height characteristic for the nanoparticles. Then, a polydimethylsiloxane (PDMS – Sylgard 184) solution was transferred onto the master to obtain a template with the opposite shape of the Si master. After 4 h of polymerization at 60° C, PDMS was peeled away and the cylindrical pillars on the template were covered with a poly(vinyl alcohol) (PVA) solution and left in the oven at 60° C for 3 h. Once polymerized, the sacrificial PVA template presented the same cylindrical wells of the original Si master.

DPNs are composed by a mixture of (poly(lactic acid-co-glycolic acid) (PLGA) and poly(ethylene glycol) diacrylate (PEG diacrylate) polymers. 30 mg of PLGA were dissolved in 600 µl of dichloromethane and chloroform, and mixed with 6 mg of PEG. Then, 0,6 mg of a photo-initiator (2-Hydroxy-4'-(2-hydroxyethoxy)-2-methylpropiophenone) were added into the polymeric solution to allow a further polymerization of PEG diacrylate. The polymeric mixture was spread into the PVA wells and the loaded templates were then exposed to UV-light for 10 min. Hydrophilic PVA templates were dissolved in deionized water for 3 h under stirring conditions at room temperature. Nanoparticles were released from the wells of PVA, and then collected through centrifugation (3,900 rpm for 20 min) and purified from PVA debris with 2 µm filters.

Purified DPNs were incubated with EDC/NHS - in a molar ratio of 3:1 (EDC/NHS:PLGA) - for 5 h under rotation at room temperature. Unlinked activators were removed with washing steps (20 min of centrifuge at maximum speed). Activated DPNs were incubated O/N with 60 µg of tPA. Unlinked drug was removed with washing steps (20 min of centrifuge at maximum speed).

**Physico-chemical Characterization of tPA-DPNs.** DPNs and tPA-DPNs size and concentration were obtained using a Multisizer 4E Coulter Particle Counter (Beckman Coulter, USA). Particles were synthetized and suspended in isotone solution. Superficial charge (ζ) was measured using Zetasizer Nano (Malvern, UK). Particles were synthetized and suspended in 1 ml of deionized water in a cuvette after shortly sonication. DPNs size and shape were observed using a Jem-1011 Transmission Electron Microscope (Jeol, Japan). DPNs were synthetized, sputtered with carbon and analyzed operating at an acceleration voltage of 100 kV. Fluorescent DPNs were synthesized adding 30 µg of Rhodamin-B (lipid-RhodB) to the polymeric mix made of PLGA and PEGDA. RhB-DPNs were activated, incubated with

FITC-tPA O/N, and suspended in PBS. Confocal images were collected with a Nikon A1 Confocal Microscope.

**Drug encapsulation efficiency.** To analyze the encapsulation efficiency, DPNs were incubated with different amount of tPA: 10, 20, 30, 40, 50, 60, and 70 µg. The amount of tPA loaded on DPNs was measured using the Bicinchoninic protein assay kit (BCA). 200 µl of Reagent were added to 25 µl of each sample of tPA-DPNs. The optical density (OD 562) was read at the micro-platereader (Tecan, CH) after 45 min of incubation at 37° C. The concentration of the drug was extrapolated by a calibration curve prepared before with different concentration of tPA dissolved in PBS.

**tPA stability.** The release profile was measured analyzing the drug still loaded on DPNs with BCA kit at different time points: 0.5, 1, 2, 4, 8, 12, 24, 48, and 72 h. In particular, particles were synthetized and incubated in PBS at 37° C. At each time point, tPA-DPNs were centrifuged to eliminate the released drug. The pellets were suspended in PBS and the optical density (OD 562) was read.

**Enzymatic Activity Test.** tPA activity was tested after the conjugation to DPNs using a chromogenic activity assay test (abcam108905). Particles were synthesized and incubated up to 2 h with the assay mix composed by plasminogen and plasmin substrate. The assay measures the ability of Tissue type Plasminogen Activator to activate the plasminogen to plasmin. The amount of plasmin produced is measured using a specific plasmin substrate releasing a yellow para-nitroaniline (pNA) chromophore.

The change in absorbance of the pNA in the reaction solution at 405 nm is directly proportional to the Tissue type Plasminogen Activator enzymatic activity.

**_In vitro_ efficacy of tPA-DPNs and tPA-SPNs in static condition.** Blood was obtained from rats after standard procedures. Anesthesia was calculated on the base of an adult rat Sprague Dawley of about 300 gr. The anesthesia was inducted with a cotton swab full of isoflurane put inside the cage. Then, 80 µl of Xilazina i.m. (right gluteus), and, after 5 min, 300 µl of Ketamine were administered. After the complete sedation of the rat, confirmed with the podal reflection, the sternum was removed and a syringe of 5 ml with a hypodermic needle of 21 G was inserted into the left ventricle to gently aspirate the blood.

Blood clots were prepared adding into several tubes 50 U of thrombin solution with 100 µl of whole blood. After 30 min of maturation at 37° C with continuous shaking at 50 rpm, clots were moved from each Eppendorf tube into a 24-multiwells dish with 1.5 ml of saline solution for each well.

Dissolution and dissolution rate of blood clots were measured estimating the amount of hemoglobin released from the clots over time. Clots were treated with: PBS, free tPA, DPNs, SPNs, tPA-DPNs, and tPA-SPNs. In particular, they were treated with 10, 30, and 50 µg of tPA/clot. Clots were incubated at 37° C with continuous shaking at 50 rpm for 300 min. At times 0, 30, 90, 180, and 300 min post treatments, 200 µl of supernatant were placed in a 96-multiwells dish and the optical density (OD 415) of the hemoglobin released in the dissolution was measured in a micro-plate reader. The dissolution rate (DR) was measured with the following formula:

$$DR = (OD_i - OD_0) / (t_i - t_0)$$

where *i* and 0 are two consecutive time points.

Moreover, to establish the possible interference of the protein corona, first, the same experiment was performed with FBS instead of PBS and, then, tPA-DPNs and tPA-SPNs were incubated for 0.5, 1, and 3 h with FBS and, then, tested to dissolve blood clots in PBS.

**Blood clot characterization.** To analyze the structural changing of blood clots before and after the treatment with tPA-DPNs, clots were fixed for 2 h in a fixative solution (2 % Glutaraldehyde, 2 % Paraformaldehyde in buffer Na-Cacodylate 0.1 M) and then post-fixed (2 h) in a solution 1 % OsO4, 1.5 % Hexacyanoferrate in Na-cacodylate buffer. Samples were then dehydrated with series of alcohols. After complete dehydration they were infiltrated with a scale of Ethanol, treated with hexametyldisilazane (HMDS) solution, left in HMDS for some hours and finally left under the hood over/night to let the evaporation of HMDS. Samples were then sputtered with 10 nm of gold.

SEM images were collected with a Jeol JSM 6490-LA (Jeol, Japan) electron microscope, operating at an acceleration voltage of 10 kV.

***In vitro* efficacy of tPA-DPNs and tPA-SPNs in dynamic condition.** The chips were fabricated by using a replica molding approach as previously described.[31] It consists in two lithographic steps (etching of the pillars and etching of the whole chip) for the creation of a silicon master that was then replicated by PDMS (Sylgard 182). The microfluidic chip consists in two micro-channels only connected *via* a 500 µm long array of pillars. The upper and lower channels have a height of 50 µm and a width of 200 µm. PDMS templates were punched to create four holes (2 inlets and 2 outlets) for the handling of the fluids. Templates were treated with $O_2$ plasma and bonded with a glass coversheet.

Blood was collected in heparin containing tubes. To create the clot, two tubes, connected to the syringe pump, were inserted in the two inlets on the microfluidic chip. One tube was filled with whole blood, and the other with thrombin solution and $CaCl_2$. Once the two solutions reached the pillars, they entered in contact and the clot started to form.

Tubes with blood and thrombin were replaced with tubes containing the different treatments: PBS, DPNs, SPNs, tPA, tPA-DPNs, and tPA-SPNs. In particular, clots were treated with 30 µg/ml of tPA. The flow rate imposed on the syringe pump was 100 nl/min. A movie of the clot dissolution was acquired at the time-lapse microscope for 2 h. The area of the clots was measured over time with ImageJ. Since the initial size of the clots was different throughout the experiments, the average of the size percentage was weighted based on the initial size of each clot. The weighted area was calculated by using the following formula:

$$\bar{x} = \frac{\sum_{i=1}^{n} A_i A_0}{\sum_{i=1}^{n} A_0}$$

**In vivo efficacy of tPA-DPNs and tPA-SPNs.** The *in vivo* study was performed in C57BL/6 mice (EJ, Le Genest, St-Berthevin, France) aged 6-7 weeks. Mice were anesthetized with intraperitoneal injection of ketamine and xylazine at 100 mg/kg and 10 mg/kg respectively. The mesentery was exposed through a midline abdominal incision and vessels were visualized by an intravital microscope (Leica MacroFluo™, Leica Microsystems SAS, Nanterre Cedex, France) using Orca Flash 4.0 scientific CMOS camera (Hamamatsu Photonics France SARL, Massy, France) to choose the vessel to use. To follow the thrombus formation, 30 μl of Rhodamine 6G (0.3 % w/v) were retro-orbitally injected. Rhodamine 6G fluorescently labels platelets and leukocytes. The chosen vessel was covered for 1-2 minutes with a 1 mm large Whatman chromatography paper soaked into a 10 % w/v iron chloride solution prepared in saline and, then, washed twice with a saline solution. The thrombus formation was monitored in real-time by fluorescence microscopy by following the accumulation of fluorescently labeled platelets (excitation 545 nm – emission 610 nm). 50 mice were randomly divided in 5 treatment groups: DPNs, SPNs, tPA 1.0 and 2.5 mg/kg, tPA-DPNs 1.0 and 2.5 mg/Kg, tPA-SPNs 1.0 and 2.5 mg/kg. Particles were retro-orbitally injected 10-15 minutes after thrombus induction. The injected volume was of 200 μL, which has been reported to be suitable to the mouse. A movie of 35 minutes was acquired. The area (A) and the intensity (I) of the clots were measured over time with ImageJ.

**Statistical analysis.** All data were processed using Excel 2010 software (Microsoft) and GraphPad PRISM. Results are expressed as mean $\pm$ standard deviation. Statistical analyses on *in-vitro* experiments were performed using ANOVA. Statistical analyses on *in-vivo* experiments were performed using t-test. The p values of <0.05 (*), <0.01 (**), and <0.001 (***) were considered to be statistically significant.

CONFLICT OF INTEREST

The authors declare no competing financial interest.

SUPPORTING INFORMATION

The Supporting Information includes additional details on the physico-chemical characterization and biological characterization of DPNs and SPNs. Movies of tPA, and DPN interaction with blood clots *in vitro* and *in vivo* are provided. Statistical analysis is also provided. The Supporting Information is available free of charge on the ACS Publications website.

**ACKNOWLEDGMENTS**

The authors wish to thank the reviewers for their valuable comments and suggestions to improve the quality of the paper. This project was partially supported by the European Research Council, under the European Union's Seventh Framework Programme (FP7/2007-2013)/ERC grant agreement no. 616695, by the Italian Association for Cancer Research (AIRC) under the individual investigator grant no. 17664, and by the European Union's Horizon 2020 research and innovation programme under the Marie Skłodowska-Curie grant agreement No 754490. The authors acknowledge the precious support provided by the Nikon Center, the Material Characterization Facility, the Electron Microscopy and Nanofabrication facilities at the Italian Institute of Technology, and the technological imaging facilities of Inserm U1148 at X. Bichat Hospital in Paris.

FIGURE CAPTIONS

**Figure.1 Physico-Chemical Properties of Discoidal Polymeric Nanoconstructs Associated with tPA Molecules (tPA-DPNs). A.** Schematic representation of tPA-DPNs, highlighting the porous structure of DPNs and their direct conjugation with tPA. **B.** Multisizer analysis of DPNs (green) and tPA-DPNs (orange). **C.** TEM image of DPNs demonstrating the circular shape with a base diameter of ∼ 1,000 nm. The upper-left inset shows a fluorescent microscopy image of a RhB-DPN superimposed on its TEM image. **D.** SEM image of DPNs demonstrating the diameter of ∼ 1,200 nm and the height of ∼ 347 nm. **E.** AFM image of DPNs demonstrating the diameter of ∼ 1,100 nm and the height of ∼ 300 nm. **F.** Size stability of tPA-DPNs in PBS at 37° C, *via* DLS analysis. [n = 3]

**Figure.2 Loading and Stability of tPA-DPNs. A.** Confocal images of DPNs co-loaded with FITC-tPA (green) and lipid-RhB (red), demonstrating the uniform distribution of both compounds throughout the porous DPN matrix. **B.** Amount of tPA associated with DPNs ($0.5 \times 10^9$) for different mass inputs of the thrombolytic agent. The dashed line represents a 1:1 ratio between associated tPA and input, demonstrating an association efficiency of almost 100%. [n = 3] **C.** tPA release from tPA-DPNs over time, at 37°C, demonstrating a stable association. [n = 3] **D.** Enzymatic activity test [n = 3]

**Figure.3 *In vitro* Efficacy of tPA-DPNs and tPA-SPNs Under Static Conditions in PBS**. **A.** Representative images of treated with PBS (CTR), free tPA (tPA), empty DPNs (DPNs), and tPA-DPNs at 5 selected time points. (tPA equivalent dose is 30 μg). **B.** Dissolution of blood clots over time, for 30 μg equivalent dose of tPA, measured *via* optical absorbance. [n = 10] **C.** Dissolution rates of blood clots at different time points, for 30 μg equivalent dose of tPA, measured *via* optical absorbance. [n =

10] **D.** TEM images of blood clots before (left) and after (right) treatment with tPA-DPNs, showing the dissolution of the fibrin network.

**Figure.4 *In vitro* Efficacy of tPA-DPNs and tPA-SPNs Under Static Conditions in FBS. A.** Representative images of blood clots treated with FBS (CTR), free tPA (tPA), empty DPNs (DPNs), and tPA-DPNs at 5 selected time points. (tPA equivalent dose is 30 µg). **B.** Dissolution of blood clots over time, for 30 µg equivalent dose of tPA, measured *via* optical absorbance. [n = 10] **C.** Dissolution rates of blood clots at different time points, for 30 µg equivalent dose of tPA, measured *via* optical absorbance. [n = 10] **D.** Dissolution of blood clots with tPA-DPNs and tPA-SPNs, pre-incubated with FBS for 30 min, 1 h and 3 h. A direct comparison is provided with fresh, free tPA. Clot dissolution is plotted over time, for 30 µg equivalent dose of tPA. [n = 10]

**Figure.5 *In vitro* Efficacy of tPA-DPNs and tPA-SPNs Under Dynamic Conditions. A.** Representative images of a blood clot forming spontaneously within the chambers of a microfluidic chip upon mixing of whole blood (bottom channel) and a thrombin solution (upper channel). **B.** Representative images of a blood clot taken at 0, 60 and 120 min post injection of free tPA. **C.** Representative images of a blood clot taken at 0, 60 and 120 min post injection of tPA-DPNs. **D.** Dissolution of the blood clot, measured as area reduction, for 30 µg equivalent dose of tPA. [n = 3] **E.** Dissolution rate of the blood clot, measured as relative area reduction over time, for 30 µg equivalent dose of tPA. [n = 3]

**Figure.6 *In vivo* Efficacy of tPA-DPNs and tPA-SPNs in a Murine Model of Thrombosis. A.** Representative images of 'beneficial' and 'critical' blood clots, occurring in a murine mesentery venule, taken at 0, 15, and 35 min post retro-orbital administration of the thrombolytic solution. **B**

Variation over time of the blood clot area (green line) and fluorescent intensity (red line) for a 'beneficial' clot. **C** Variation over time of the blood clot area (green line) and fluorescent intensity (red line) for a 'critical' clot. **D.** Percentage occurrence of 'beneficial' (in blue) and 'critical' (in red) blood clots at 35 min post treatment initiation for seven different experimental groups – empty DPNs (DPNs); empty SPNs (SPNs); 1.0 mg/kg of free tPA (tPA 1); 2.5 mg/kg of free tPA (tPA 2.5); 1.0 mg/kg of tPA-SPNs (tPA-SPNs 1); 2.5 mg/kg of tPA-SPNs (tPA-SPNs 2.5); 1.0 mg/kg of tPA-DPNs (tPA-DPNs 1); 2.5 mg/kg of tPA-DPNs (tPA-DPNs 2.5). [n ≥ 10 mice] **E.** Percentage of blood clot area, average weighed on the initial area, at 35 min post treatment initiation for five different experimental groups – empty DPNs (DPNs); empty SPNs (SPNs); 1.0 mg/kg of free tPA (tPA 1); 2.5 mg/kg of free tPA (tPA 2.5); 1.0 mg/kg of tPA-SPNs (tPA-SPNs 1); 2.5 mg/kg of tPA-SPNs (tPA-SPNs 2.5); 1.0 mg/kg of tPA-DPNs (tPA-DPNs 1); 2.5 mg/kg of tPA-DPNs (tPA-DPNs 2.5). [n ≥ 10 mice]

FOR TABLE OF CONTENTS ONLY

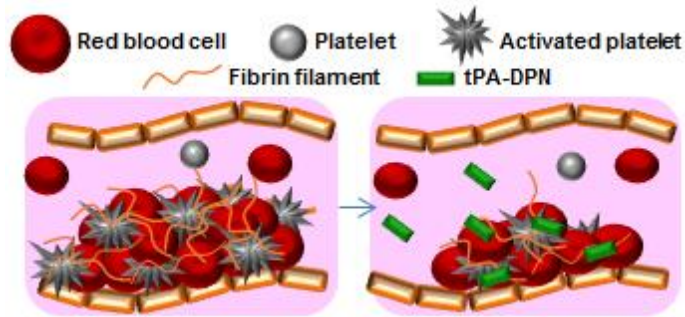

Tissue plasminogen activator-conjugated discoidal polymeric nanoconstructs (tPA-DPNs) enhance the thrombolytic activity, extent the circulation half-life and prevent the rapid degradation of tPA.